\documentclass[aps,prl,unsortedaddress, %superscriptaddress,
               showpacs,footinbib,
               amsfonts,amssymb,amsmath,nobibnotes,
               twocolumn, reprint,  % <--- For final version
              ]{revtex4-1}

\usepackage[utf8]{inputenc}
\usepackage{tabularx}
\usepackage{multirow}
\usepackage{graphicx}
\usepackage{color}
\usepackage{mathrsfs}
\usepackage{bm}  % bold greek letters
\usepackage{braket}
\usepackage{placeins}
\usepackage{pbox}
\usepackage{lipsum}

\usepackage [autostyle, english = american]{csquotes}
\MakeOuterQuote{"}

\usepackage[breaklinks]{hyperref}

\usepackage[hyphenbreaks]{breakurl}

\usepackage{fancyhdr}

\graphicspath{{img/}{img/bandstructures/}{img/absorption/}
              {img/excitons/}{img/wavefunctions/}{img/other/}
              {img/cartoons/}}

\newlength{\imwidth}
\newlength{\imheight}
\newlength{\mylabelwidth}

\newcolumntype{i}{>{\centering\arraybackslash}m{\imwidth}}
\newcolumntype{I}[1]{>{\centering\arraybackslash}m{#1}}
\newcolumntype{a}{m{\mylabelwidth}}{}
\newcolumntype{L}{>{$}l<{$}}{}
\newcolumntype{R}{>{$}r<{$}}{}

\newcommand{\berkeley}{%
  Department of Physics, University of California at Berkeley,
  California 94720, USA
  and Materials Sciences Division, Lawrence Berkeley National Laboratory,
  Berkeley, California 94720, USA
  }

\newcommand{\berkeleymail}{antonius@lbl.gov}

\begin{document}

\title{Orbital symmetry and the optical response of single-layer MX monochalcogenides}
\author{Gabriel Antonius} \email{\berkeleymail}
\author{Diana Y. Qiu} \email{diana.qiu@berkeley.edu}
\author{Steven G. Louie} \email{sglouie@berkeley.edu} %\phone{(510) 642-1709}

\affiliation{\berkeley}

\begin{abstract}

We show that the absorption spectra of single-layer GaSe and GaTe in the
  hexagonal phase feature exciton peaks with distinct polarization selectivity.
We investigate these distinct features from first-principles calculations
  using the GW-BSE formalism.
We show that, due to the symmetry of the bands under in-plane mirror symmetry,
  the bound exciton states selectively couple to either
  in-plane or out-of-plane polarization of the light.
In particular, for a $p$-polarized light absorption experiment,
  the absorption peaks of the hydrogenic $s$-like excitons emerge
  at large angle of incidence,
  while the overall absorbance reduces over the rest of the spectrum.
 
\end{abstract}

\maketitle

Two-dimensional (2D) materials receive continued interest
  as building blocks for nanoscale electronics.
The ability to thin down a material to atomically thin layers
  allows for a finer electrostatic control over the system
  \cite{radisavljevic_integrated_2011, radisavljevic_single-layer_2011}
  and opens for the possibility of designing layered heterostructures
  \cite{geim_van_2013}.
Beyond graphene, 2D materials with a sizeable band gap
  and good electrical conductivity find applications
  in field-effect transistors
  and optoelectronic devices \cite{wang_electronics_2012}.
The band gap of these materials changes dramatically
  with reduced number of layers due to quantum confinement effects
  and reduced screening
  \cite{mak_atomically_2010,Bradley2015,Li2017},
  and single-layer materials show an enhanced modulation of the band gap
  with applied strain~\cite{shi_quasiparticle_2013}.

A large body of research focuses on transition metal dichalcogenides
  $MX_2$ (e.g. M=Mo,W; X=S,Se) due to their novel optical features, such as
  strongly bound excitons and trions~\cite{Ashwin2012, Qiu2013, Mak2013,
    Chernikov2014, Ugeda2014, Klots2014, Zhu2014,
    Ye2014,Hill2015,Qiu2015,qiu_screening_2016}, 
  valley-selective circular dichroism and coupling of spin and valey degrees
  of freedom~\cite{Cao2012,Xiao2012,Mak2012},
  and large optical absorbance by single monolayers~%
  \cite{mak_atomically_2010}.
Likewise, the metal monochalcogenides $MX$ (e.g. M=Ga,In; X=S,Se,Te)
  arouse interest for their electrical and optical properties
  \cite{yang_highly_2016, xu_synthesis_2016, xiong_one-step_2016,
        qasrawi_optical_2016, jung_red-ultraviolet_2015,
        huang_highly_2015, cao_strong_2015, zhou_epitaxy_2014,
        li_controlled_2014, lei_synthesis_2013, late_gas_2012,
        hu_synthesis_2012}.
In particular, gallium selenide was shown to have exceptionally large
  photoresponse~\cite{huang_highly_2015},
  and strong second harmonic generation \cite{jie_layer-dependent_2015}.
It has been successfully used in designing photodetectors
  and phototransistor devices 
  by means of chemical vapor deposition or pulsed laser deposition
  \cite{mahjouri-samani_pulsed_2014, li_controlled_2014,
        hu_synthesis_2012, cao_strong_2015}.
Gallium telluride also exhibits a large photoresponsitivity
  and has a smaller band gap than GaSe,
  making it suitable for photodetector applications
  \cite{wang_high-performance_2015}.

So far, all devices made with GaSe and GaTe used multi-layers samples,
  with very few measurements having been performed on single-layer samples.
Those measurements performed on multi-layer GaSe show a strong
  anisotropy in the optical absorption spectrum,
  likely related to the selection rules
  of the single layer \cite{liang_optical_1975}.
The single-layer materials may however exhibits distinct properties
  from their bulk counterpart.
For example, upon decreasing the number of layers,
  the band structure of GaSe undergoes a direct-to-indirect
  band gap transition \cite{li_controlled_2014,rybkovskiy_transition_2014}.
As a result, single-layer GaSe exhibits a sharp peak in the density of states
  near the Fermi level \cite{cao_tunable_2015},
  potentially enhancing optical absortion in this material.
The optical absorption may be further enhanced by excitonic effects,
  which are known to be large in low dimensional systems
  due to enhanced electron-electron interaction
  and reduced screening~\cite{Spataru2004,Wang2005,Deslippe2009,Qiu2013,
                              Chernikov2014,qiu_screening_2016}.

In this work, we present a first-principles study
  of the optical absorption spectrum
  of hexagonal single-layer GaSe and GaTe.
We obtain the quasiparticle bandstructure from the {\it ab initio}
  GW formalism~\cite{Hedin1965,Hybertsen1986}
  and, with the GW results, compute the optical properties by
  solving the Bethe-Salpeter equation (BSE)~\cite{Strinati1988,Rohlfing2000}.
We show that the exciton states in these single-layer MX monochalcogenide
  materials acquire selection rules
  for linear optical excitations coming from the symmetry of the band states
  under in-plane mirror reflection,
  similar to observations in some layered materials~\cite{%
  ArnaudHugeExcitonicEffects2006,
  Brotons-GisbertNanotexturingEnhancePhotoluminescent2016}.
The optical properties of the monolayers
  are thus strongly modulated with the angle of incident light
  with respect to the normal direction of the plane.
We find that the binding energy of the lowest energy exciton                    
  in GaSe is 0.66 eV, which is comparable to the large excition binding         
  energies ranging between 0.3 and 0.7 eV
  in transition metal dichalcogneides~%
  \cite{Qiu2013,Mak2013,Ugeda2014,Chernikov2014,Klots2014,Ye2014,Zhu2014}.

\section{Structural parameters and GW-BSE calculation}

For the ground-state properties,
  we performed density functional theory (DFT) calculations
  with the Quantum Espresso software \cite{giannozzi_quantum_2009}
  and the Abinit software \cite{Gonze2009}.
We use norm-conserving pseudopotentials with a kinetic energy cutoff
  of $250$~Ry and include the semi-core $n=3$ shell of gallium as part of
  the valence electrons.
For gallium selenide, 
  we use a hexagonal unit cell,
  which is the structural phase found
  for both the monolayer and the bulk crystals.
The lattice parameter of GaSe ($3.75 \AA$) was taken from experiments
  \cite{li_controlled_2014}
  and the internal degrees of freedom were relaxed to minimise the forces
  using an LDA exchange-correlation functional.
Unlike gallium selenide, gallium telluride forms a monoclinic structure
  in the bulk crystal \cite{wang_high-performance_2015}.
Our calculations, however, indicate that the hexagonal phase
  of the single-layer GaTe is more stable than the monoclinic one
  (by only $4$~meV/atom).
Therefore, we use the same hexagonal phase for both GaSe and GaTe.
The structural parameters of GaTe were optimized
  with a PBE~\cite{Perdew1996} exchange-correlation functional,
  giving a lattice parameter of $4.14~\AA$.
The use of LDA vs. PBE does not change the symmetry of the band states,
  and only changes the Kohn-Sham gap by $\sim 0.1$~eV which is small
  compared to the self-energy correction to the gap (which is $\sim 1.3$~eV).

The GW-BSE calculations were performed with the BerkeleyGW software
  \cite{Hybertsen1986,Rohlfing2000,deslippe_berkeleygw_2012}. 
We used a k-point sampling of $12\times12\times1$,
  a kinetic energy cutoff of $30$~Ry for the dielectric matrix,
  and 600 bands to compute the dielectric matrix and the self energy.
We add a static remainder correction to the self energy
  to accelerate convergence with respect to the summation
  over empty states~\cite{Deslippe2013}.
We included 14~$\AA$ of vacuum in the out-of-plane direction and truncated
  the Coulomb interaction in this direction to prevent spurious
  interactions between periodic images~\cite{Ismail-Beigi2006}.
The frequency dependence of the dielectric matrix was included within the
  Hybertsen-Louie generalized plasmon pole model (HL-GPP)~\cite{Hybertsen1986}.
Achieving a sufficiently dense k-point sampling for the BSE calculation
  is especially challenging in 2D materials,
  due to the fast spatial variation of the screening
  \cite{Ismail-Beigi2006,Cudazzo2011,huser_how_2013,Qiu2013,%
        Wirtz2013,Chernikov2014,qiu_screening_2016,Rasmussen2016}. 
In order to accelerate the convergence of the k-point sampling
  while maintaining a reasonable computational cost,
  we interpolate the BSE kernel onto a uniform $64\times64\times1$ k-point grid
  and solve the BSE on this fine grid \cite{rohlfing_electron-hole_1998}.
The detailed structure of the screening on the fine grid
  is further refined using the clustered sampling interpolation (CSI)
  technique developed specifically for 2D materials~%
  \cite{qiu_screening_2016,Jornada2017}.
10 clustered points were used, which is equivalent to sampling the dielectric
  matrix using $(120)^2$ $q$ points on a uniform grid.
The basis set used to describe the excitons includes five valence bands
  and two conduction bands.
We verified that the absorption spectrum
  is converged with respect to the number of bands % valence and conduction bands.
  in the energy range from zero up to $5.5$~eV.
We neglect spin-orbit interactions throughout this calculation,
  in order to reduce the size of the two-particle basis
  and keep the solution of the BSE computationaly manageable.

\section{Results and discussion}

\begin{figure}
  \includegraphics[width=0.495\linewidth]{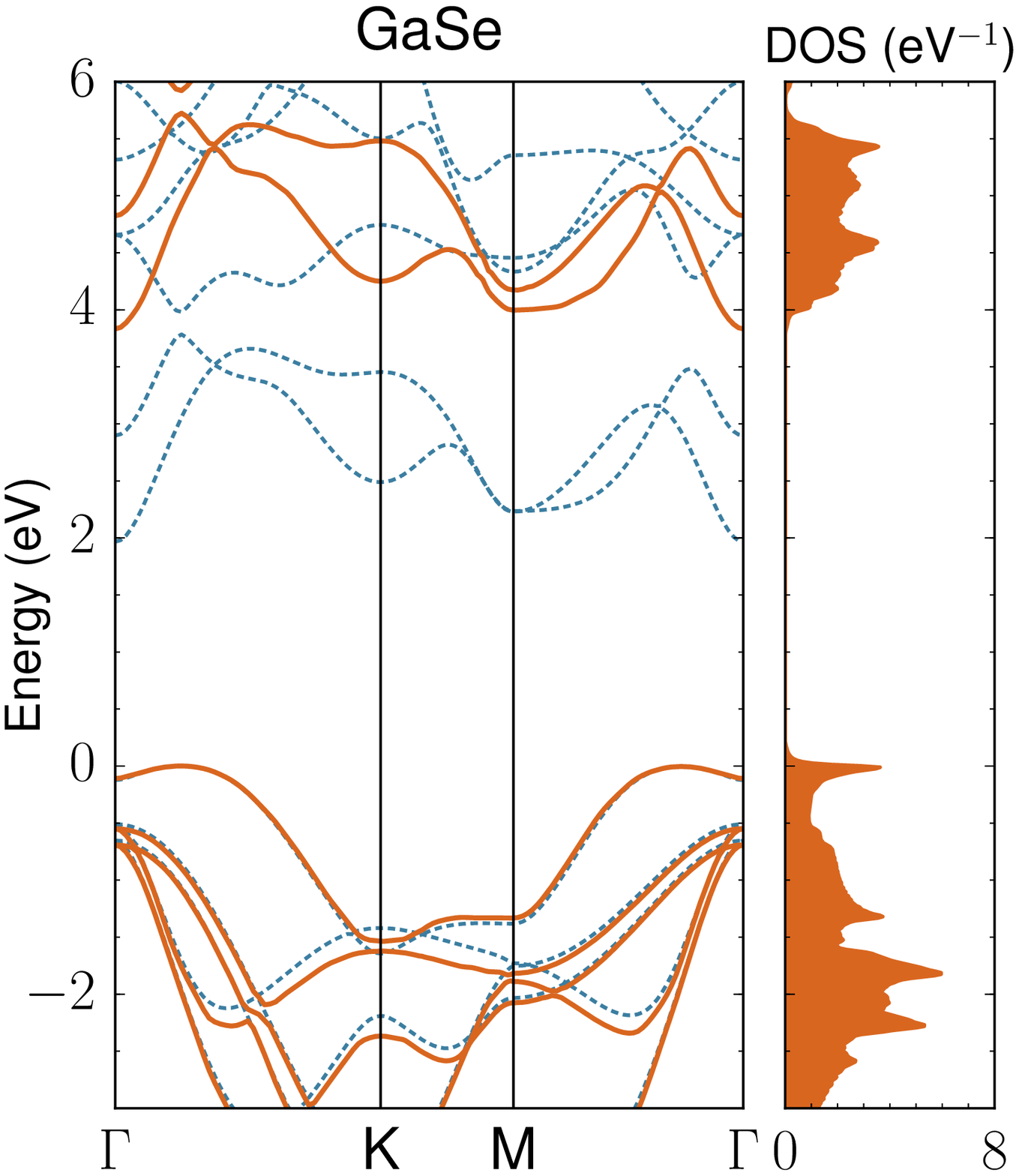}%
  \hfill
  \includegraphics[width=0.495\linewidth]{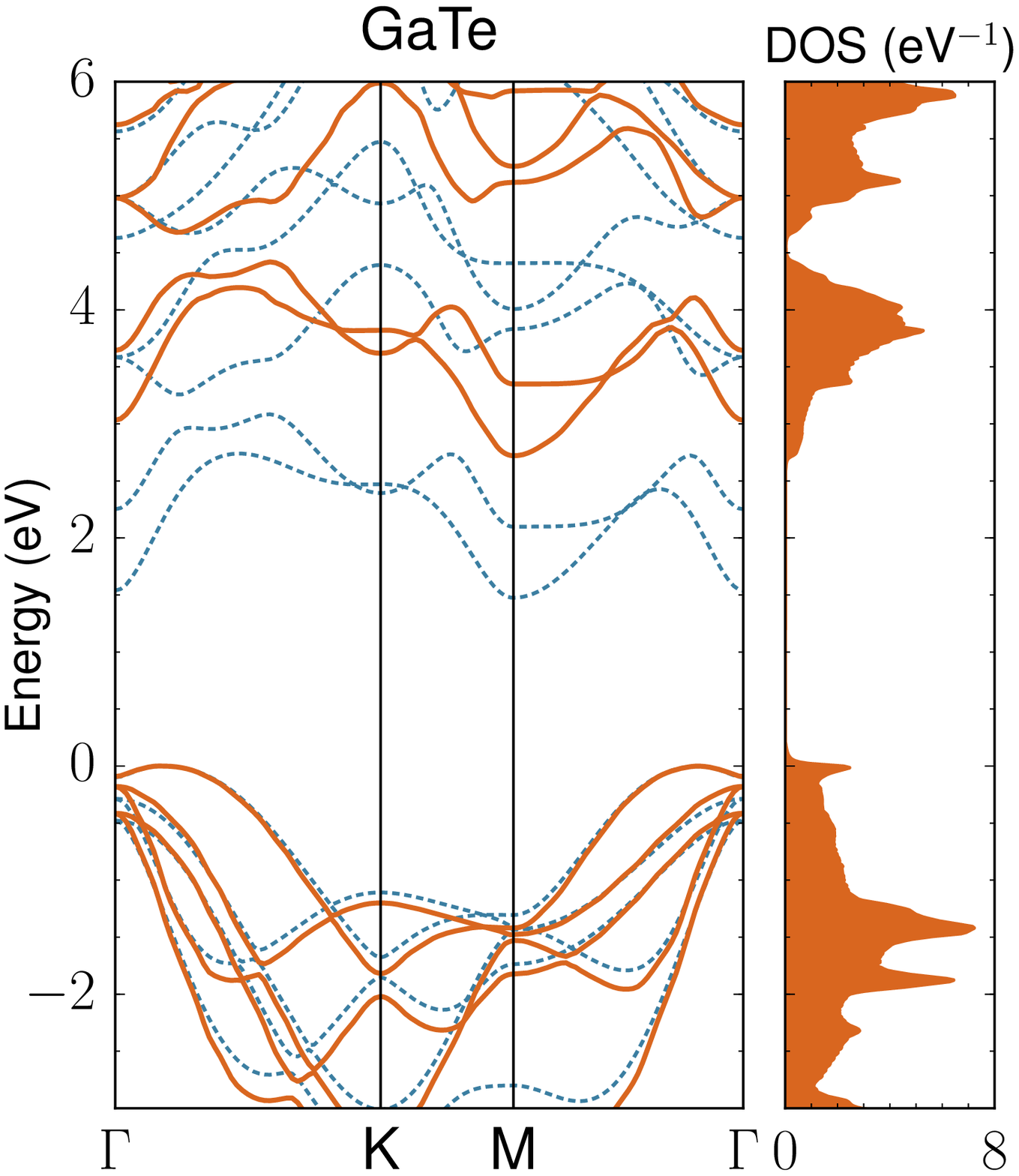}
  \caption{\label{fig:bandstructure}
    Band structures and density of states of single-layer
    GaSe (Left) and GaTe (right).
    The dashed blue lines are Kohn-Sham eigenvalues computed from DFT,
    and the solid orange lines are quasiparticle energies
    obtained within the $G_0W_0$ approximation.
  }
\end{figure}

Figure \ref{fig:bandstructure} shows the band structure and density of states
  of GaSe computed from G$_0$W$_0$ on top of DFT calculation. 
The G$_0$W$_0$ corrections increase the direct band gap of GaSe
  at $\Gamma$ from $2.10$~eV to $3.94$~eV,
  and the direct band gap of GaTe at $\Gamma$ from $1.81$~eV to $3.12$~eV.
The spin-orbit interaction (not included in our GW-BSE calculation)
  would reduce the direct band gap of GaSe by 0.05~eV,
  and that of GaTe by 0.17~eV.
The valence band maximum is located along the $\Gamma$\textendash$K$ line
  for both materials, while the conduction band minimum
  is located at $\Gamma$ in GaSe and at $M$ in GaTe.
Both material feature a caldera shape for the highest occupied band
  around $\Gamma$, with saddle points located along
  the $\Gamma$\textendash$M$ lines.
As a result, a van Hove singularity appears in the density of states
  near the Fermi level.

\begin{figure}

  \begin{center}

    \begin{tabularx}{\textwidth}{  i L | R i }% \hline
      &
      &  \Gamma^{+}_{c}
      &  \includegraphics[width=\imwidth]{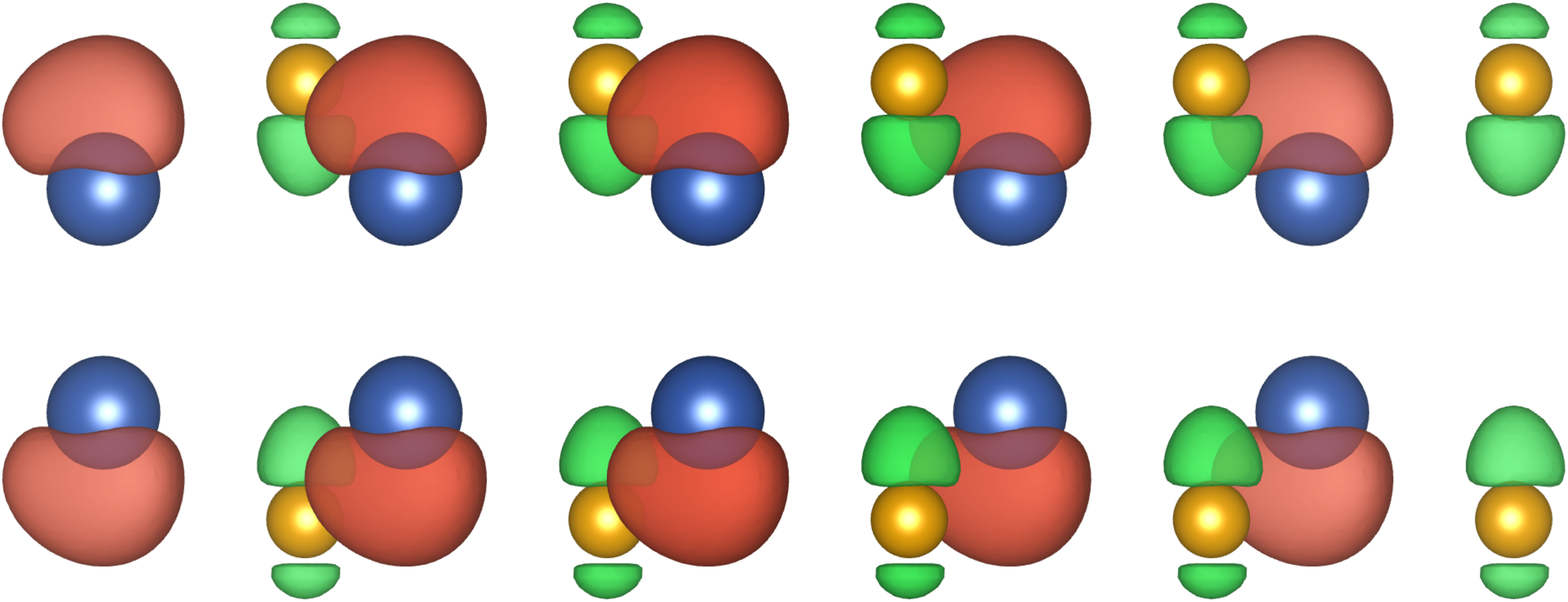}\\
      &
      &  \Gamma^{-}_{c}
      &  \includegraphics[width=\imwidth]{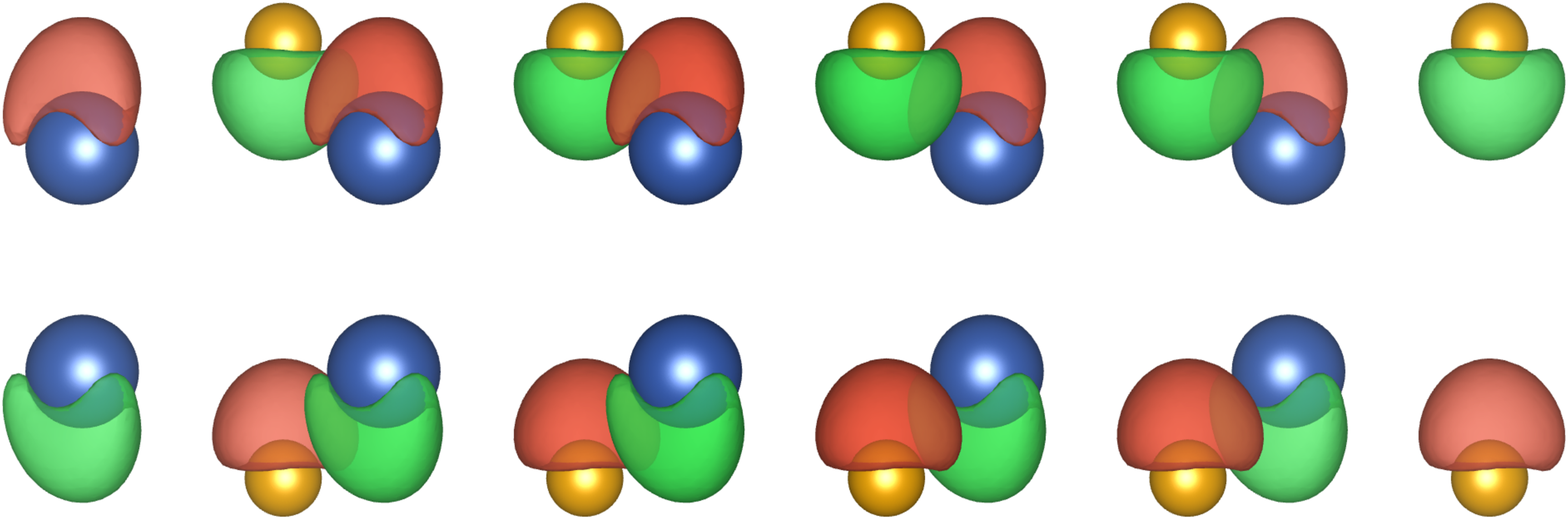}\\
         \includegraphics[width=\imwidth]{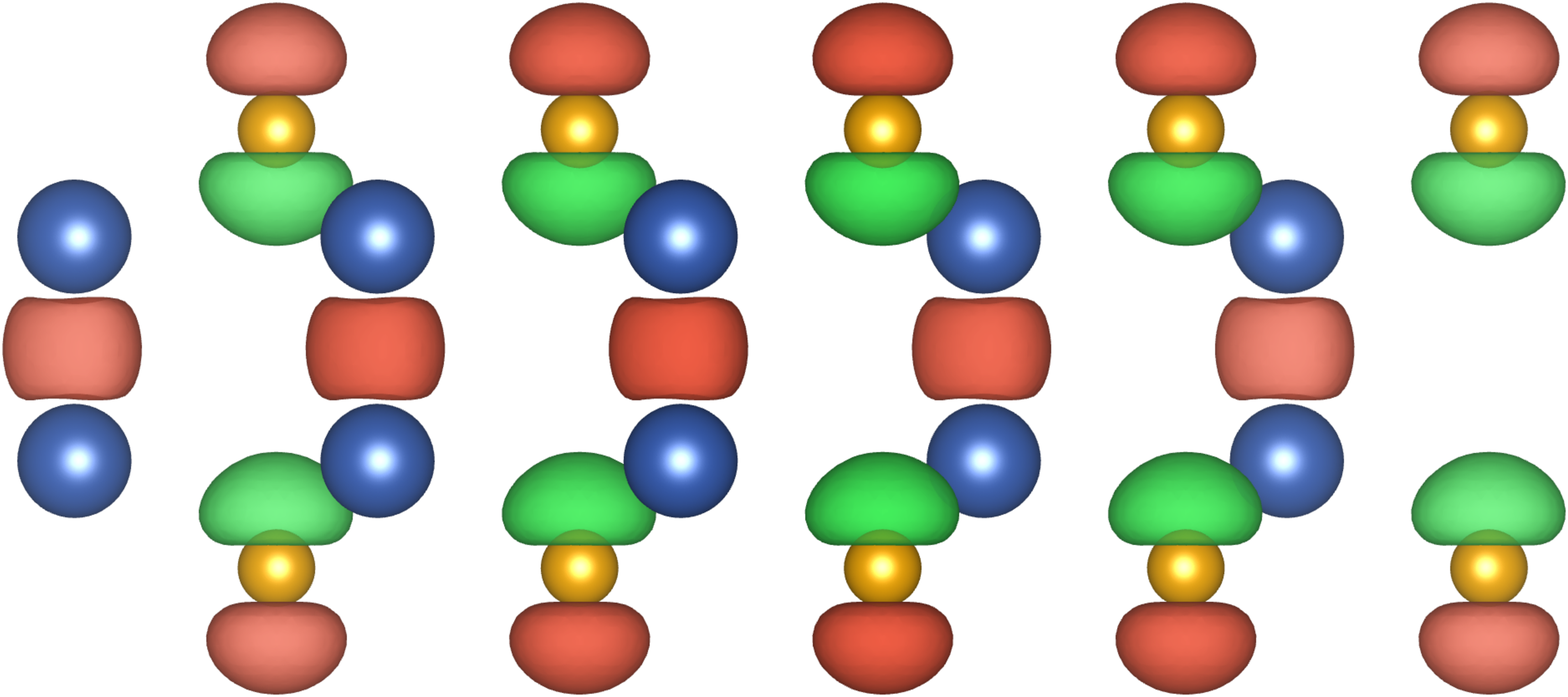}
      &  \Gamma^{+}_{v1}
      &\\
         \includegraphics[width=\imwidth]{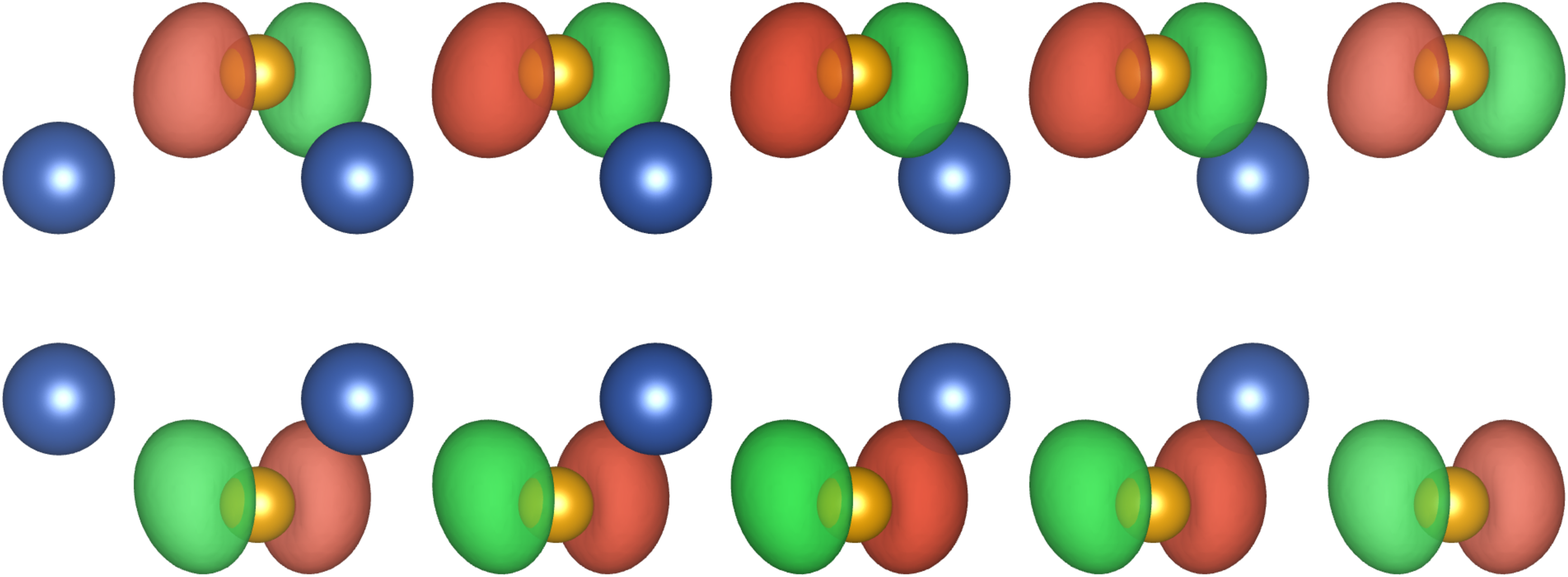}
      &  \Gamma^{-}_{v2}
      &  \\
         \includegraphics[width=\imwidth]{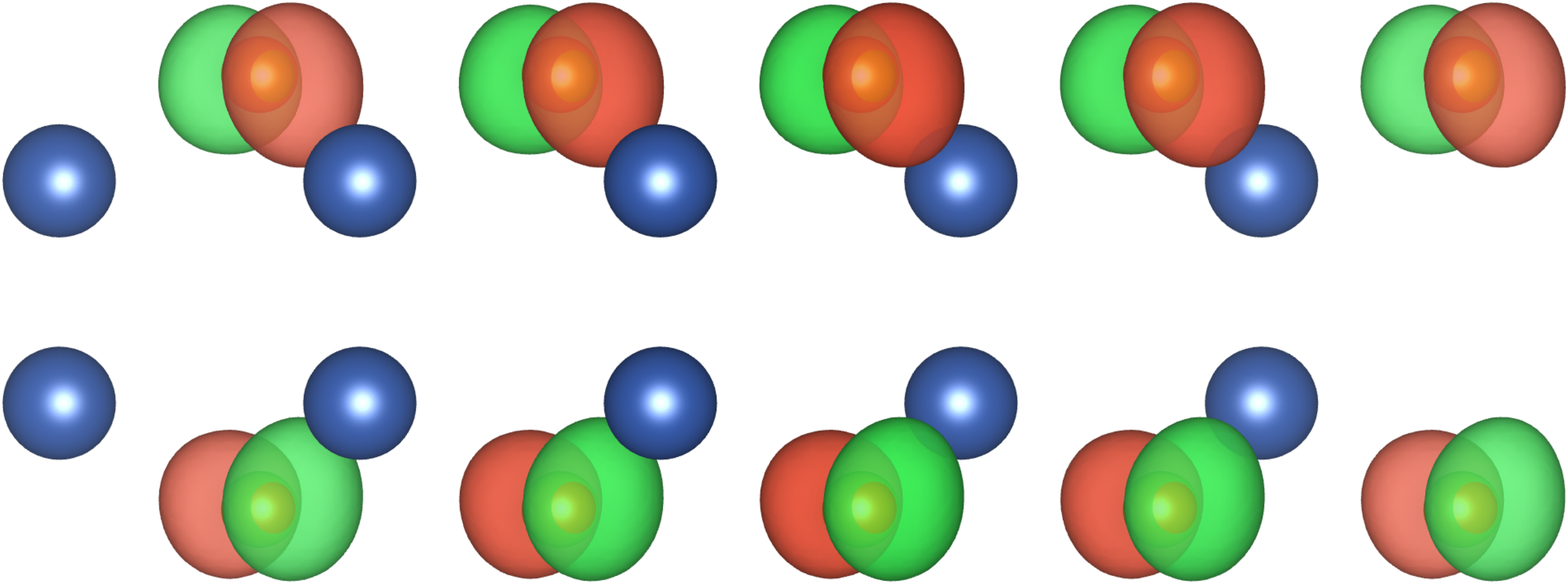}
      &  \Gamma^{-}_{v3}
      &\\
    \end{tabularx}

  \end{center}

  \caption{\label{fig:wfn}
    Electronic wavefunctions of single-layer GaSe at $\Gamma$
    (side view, with the blue balls denoting the Ga atoms
    and the yellow balls denoting the Se atoms).
    The three valence bands and two conduction bands
    are ordered in ascending energy from bottom to top.
    }
\end{figure}

\begin{table}[t]
  \begin{tabular}{c | r r r r r}
 & $\quad v_3 $  & $\quad v_2 $  & $\quad v_1 $ 
 & $\quad c_1 $  & $\quad c_2 $\\[5pt]
\hline
  $\quad \Gamma \quad$  & $-1$ & $-1$ & $ 1$ & $-1$ & $ 1$\\
  $\quad K \quad$       & $ 1$ & $-1$ & $ 1$ & $ 1$ & $-1$\\
  $\quad M \quad$       & $-1$ & $-1$ & $ 1$ & $ 1$ & $-1$\\
  \end{tabular}
  \caption{
  Parity of the electronic states in single-layer GaSe
  under in-plane mirror symmetry
  $\sigma_h$ at different points of the Brillouin zone.
  \label{tab:Zsym}}
\end{table}

It is instructive to look at the bands
  involved in the lowest optical transitions.
Figure~\ref{fig:wfn} shows the electronic wavefunctions in GaSe
  for the last three valence bands and the first two conduction bands
  at $\Gamma$.
The symmetry analysis of the bands in single-layer GaSe has been reported
  \cite{LiSymmetrydistortedband2015a},
  and here we focus on the parity of the bands
  under in-plane mirror reflection $\sigma_h$. %bisecting the z-axis.
Since the mirror reflection takes $z$ to $-z$
  (where $z$ is along the out-of-plane direction),
  the band-state wavefunctions are eigenstates of $\sigma_h$
  with even ($+$) or odd ($-$) parity.
The parity of the bands at high-symmetry points of the Brillouin zone
  is listed in Table~\ref{tab:Zsym}.
Note that the $\sigma_h$ parity is not restricted to high-symmetry points;
  in the present spin-unpolarized calculation,
  it is a good quantum number everywhere in the 2D Brillouin zone.
The first two conduction bands have opposite parities
  and exchange order in various regions of the Brillouin zone.
The crossing of these two bands can be seen in Figure~\ref{fig:bandstructure}
  along the $\Gamma-K$ and $\Gamma-M$ lines.

\begin{figure}
  \includegraphics[width=9cm]{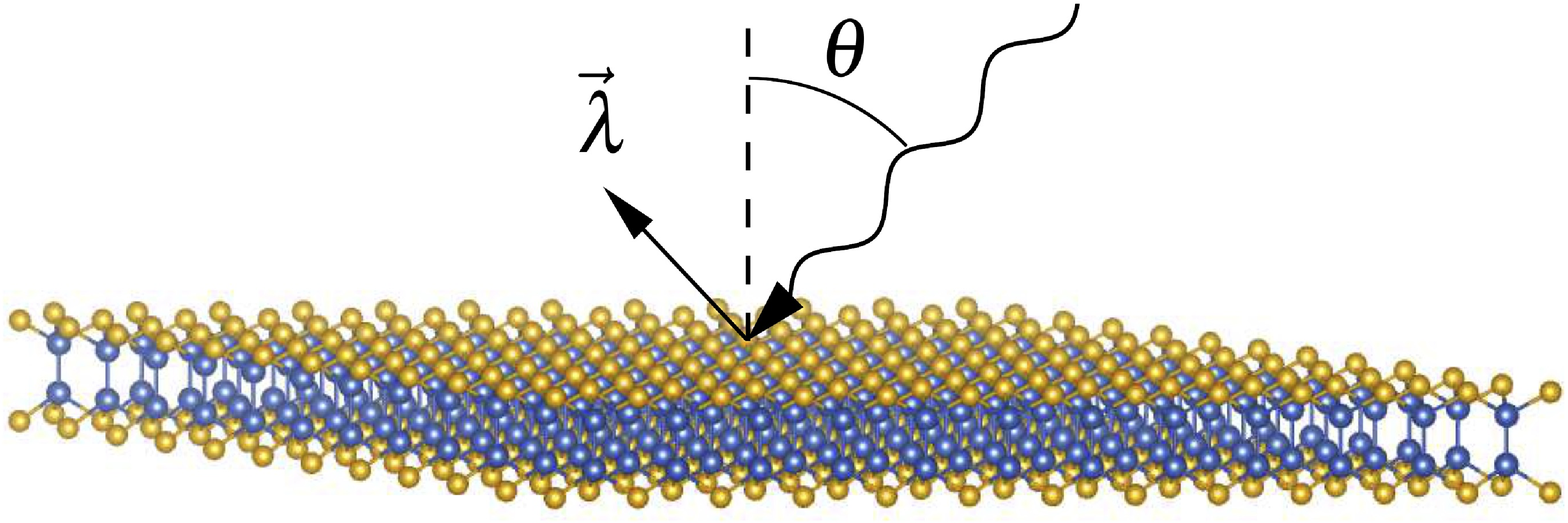}
  \includegraphics[width=9cm]{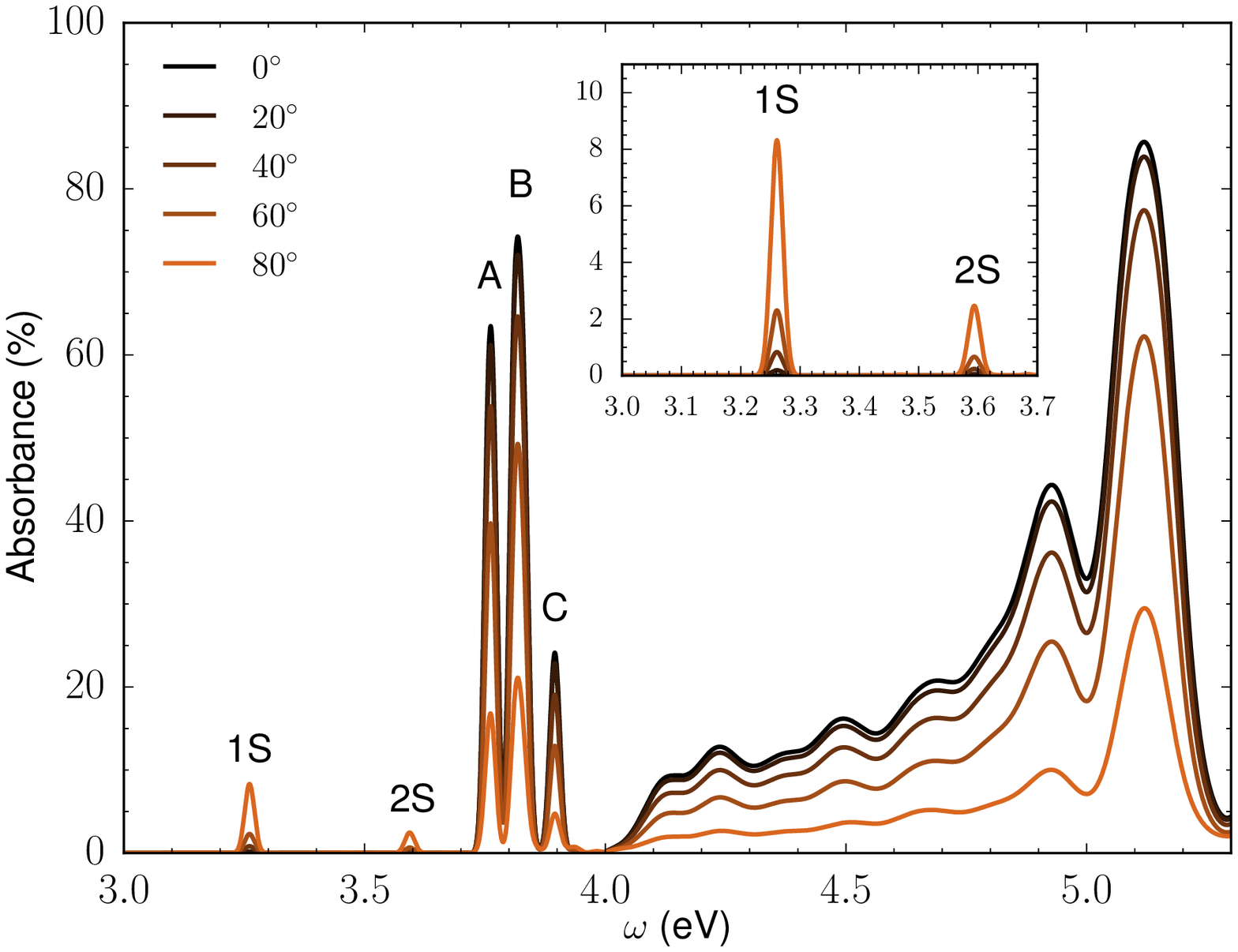}
  \includegraphics[width=9cm]{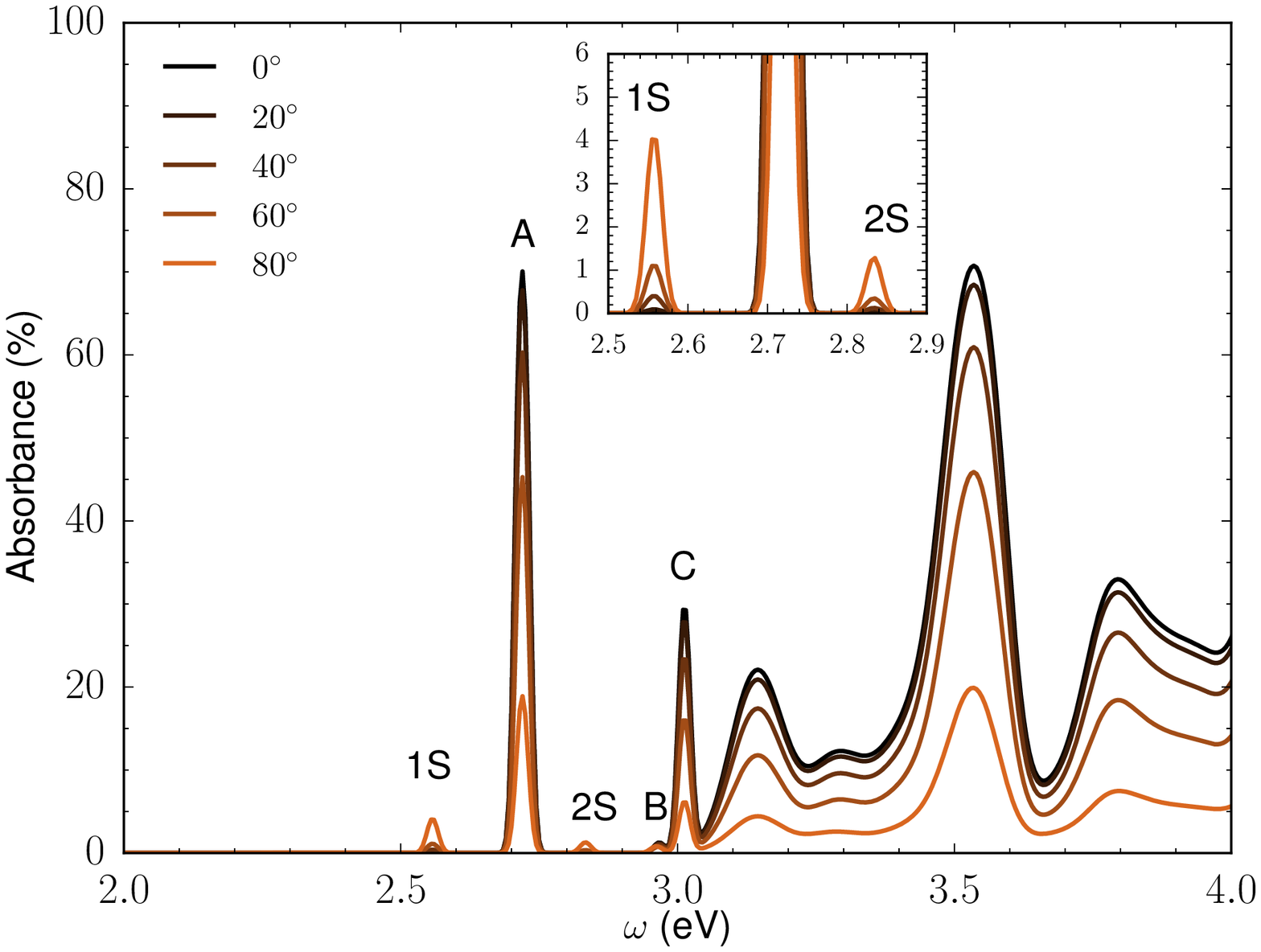}
  \caption{\label{fig:absorption-angle}
    Absorption spectrum of single-layer
    GaSe (middle) and GaTe (bottom)
    as a function of the light's angle of incidence for a p-polarized experiment,
    as depicted in the top panel.
    The insets offer a zoom on the $1s$ and $2s$ excitons'
    absorption peaks.
  }
\end{figure}

\begin{figure*}[t]

  \begin{center}

  \begin{tabularx}{\textwidth}{ m{0.13\textwidth} m{0.84\textwidth}}
    \includegraphics[width=0.13\textwidth]{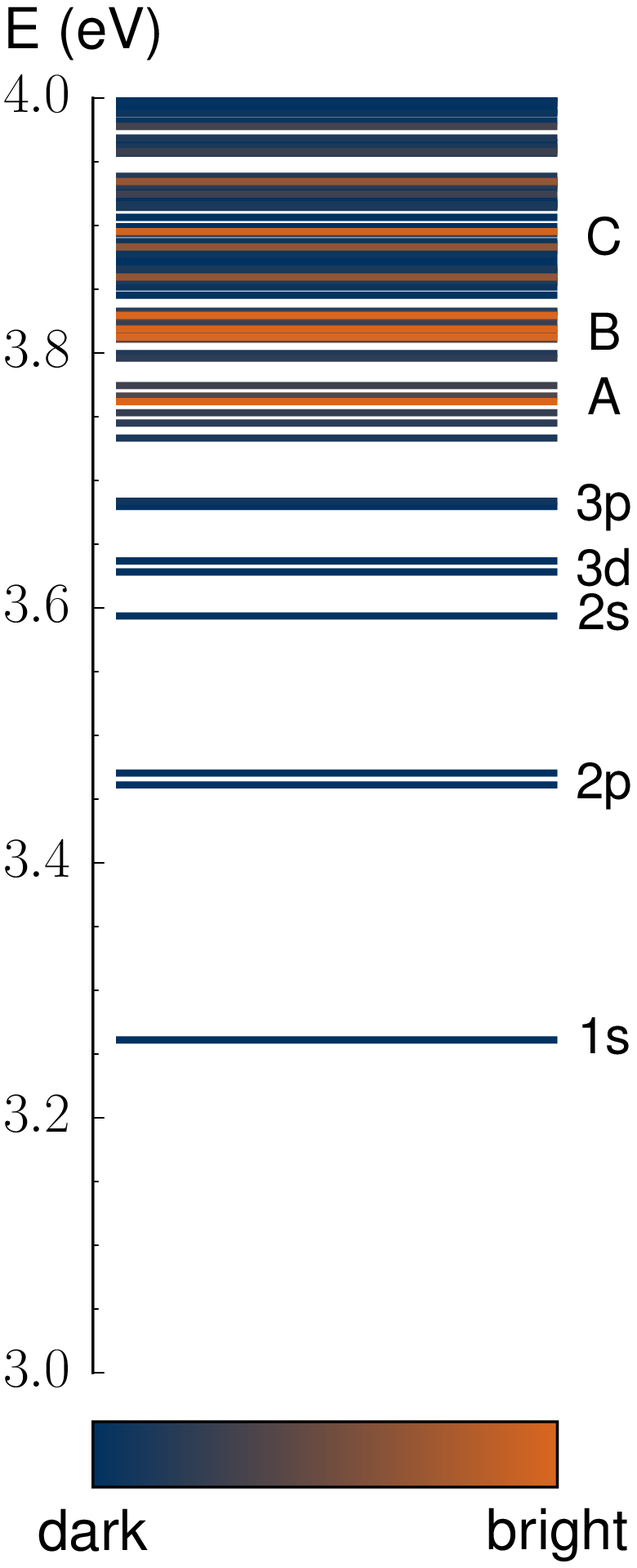}
    &
    \includegraphics[width=0.80\textwidth]{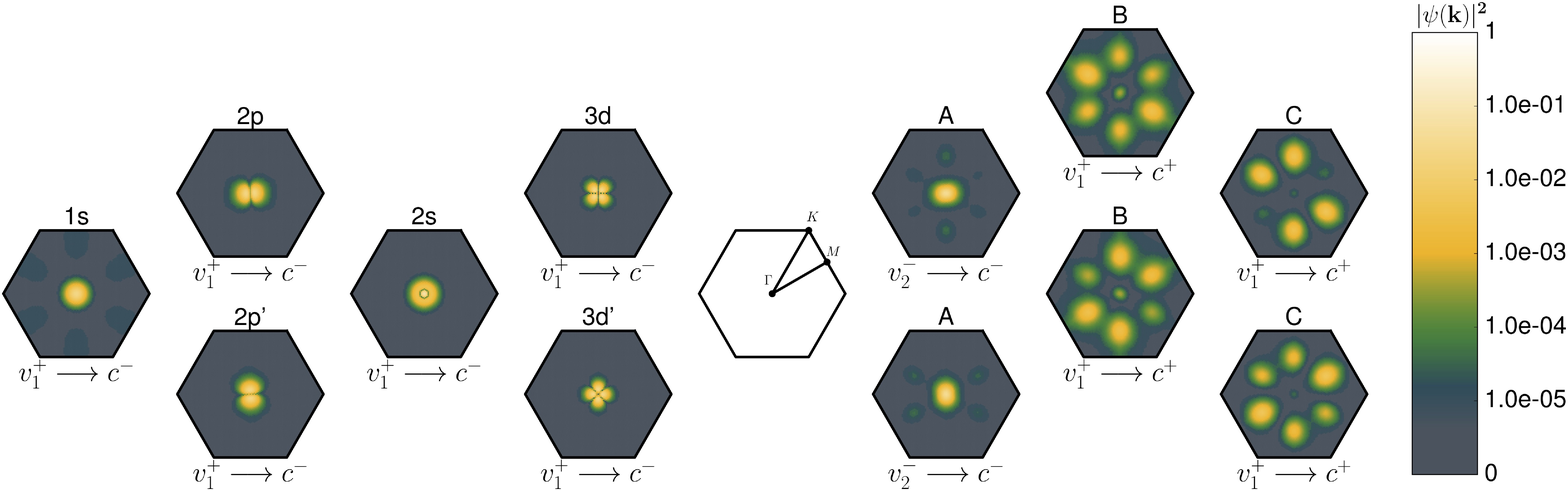}
  \end{tabularx}
  \end{center}
  \caption{\label{fig:xct}
    (Left) Energy spectrum of the bound excitons in GaSe.
    The bandgap, and hence the continuum edge, is at 3.94~eV.
    The colors indicate the squared amplitude of the dipole transition
    matrix elements (arbitrary units),
    giving the brightness of the corresponding exciton
    for in-plane polarization of the light.
    (Right) Exciton wavefunction in reciprocal space for the first few excitons
    of the hydrogenic series and the bright bound excitons of GaSe.
    At each k-point, the square norm of the electron-hole coefficient
    forming the exciton wavefunction is summed over all valence-conduction
    band pairs. The most important valence-to-conduction band transition
    is indicated below each exciton wavefunction.
  }
\end{figure*}

We compute the absorption spectrum of single-layer GaSe and GaTe
  for a $p$-polarized light absorption experiment.
The polarization vector of light is taken to lie in the plane of incidence
  oriented along the $\Gamma$\textendash$M$ direction.
Three bright exciton peaks are easily identified,
  located below the fundamental band gap,
  as can be seen in Figure~\ref{fig:absorption-angle}.
These are bound excitons, labeled A, B, and C.
They are followed by a continuum of states available for absorption above
  the fundamental band gap.

Over most of the spectrum, the absorbance decreases
  with larger angles of incidence
  (as defined in the top panel of Fig.~\ref{fig:absorption-angle})
  due to a local field or depolarization field effect.
This effect arises from the finite thickness of the layer,
  allowing for charges to accumulate on each side of the layer
  in response to a perpendicular electric field.
The depolarization field thus created reduces significantly the absorbance
  for the out-of-plane component of the light polarization
  of the applied electromagnetic wave.

A few new absorption peaks however \emph{emerge}
  within the quasiparticle gap
  with larger light angles of incidence.
These peaks are located at $3.26$~eV and $3.60$~eV in GaSe,
  and at $2.56$~eV and $2.83$~eV in GaTe.
They are associated with a lower energy series of excitons.
We label these as $1s$ and $2s$ excitons following
  the notation of the hydrogenic model
  based on the nodal structure of the envelope of the exciton wave function.

Figure~\ref{fig:xct} shows the energy levels of the bound excitons
  of GaSe along with the $k$-space distribution of the squared amplitude
  of the free electron-hole pairs (from solving the Bethe-Salpeter equation)
  that form the exciton wavefunctions in the Brillouin zone.
Similarly to other 2D semiconductors, the exciton levels do not follow
  the Rydberg series for the the 2D hygrogenic model due to spatially varying
  screening~\cite{Qiu2013,Chernikov2014,qiu_screening_2016,
                  QiuEnvironmentalScreeningEffects2017}.
Nonetheless, from the nodal structure of the wavefunctions,
  we can identify for the few lowest energy states
  a hydrogen-like series of states with clear angular moment assignment.
This series arises from the parabolic dispersion
  of the first band-to-band transition near $\Gamma$.

We define the binding energy of an exciton as the difference between
  the exciton's energy, and the energy of the dominant band-to-band
  transition at $\Gamma$, where the lowest energy direct transition occurs
  for the bands of interest.
In single-layer GaSe, this gives binding energies of $0.66$~eV and $0.34$~eV
  for the $1s$ and $2s$ excitons, respectively,
  and binding energies of $0.60$~eV, $0.13$~eV, and $0.07$~eV
  for the A, B, and C excitons, respectively.
We note, however, that the maximum amplitude of the B and C excitons
  is not located at $\Gamma$,
  but near the six-fold degenerate saddle points along the $\Gamma-M$ lines.
The direct band gap at these saddle points is $0.88$~eV larger than at $\Gamma$,
  explaining the seemingly small binding energy of the B and C excitons
  in comparison with the A exciton.

As shown in the energy diagram of Figure~\ref{fig:xct},
  the full "hydrogenic" series is dark
  under illumination with in-plane polarization of the light.
For the $p$ and $d$ excitons,
  this is due to the destructive interference of the interband
  dipole matrix elements around $\Gamma$
  for dipole-allowed interband transition systems~\cite{qiu_screening_2016}.
In contrast, the brightness of the $s$ excitons
  depends on the polarization angle of the light
  for the p-polarized experimental setup in Fig.~\ref{fig:absorption-angle}.
This selection rule originates from the symmetry
  of the quasiparticle band states under reflection along the $z$ axis.

Since the Coulomb interaction is even under in-plane mirror reflection,
  it follows that the $\sigma_h$ parity is also a good quantum number
  for all the exciton states.
The excitons with even $\sigma_h$ parity, such as the A, B, and C excitons,
  are composed of electronic transitions between band states
  with the same $\sigma_h$ parity,
  while the exctions with odd $\sigma_h$ parity,
  such as those of the hydrogenic series, are composed of electronic
  transitions between band states with opposite $\sigma_h$ parities.
For in-plane polarization of the light,
  the excitons with even parity will be bright
  and those with odd parity will be dark.
Conversely, for out-of-plane polarization of the light,
  the excitons with even parity will be dark
  and those with odd parity will be bright.

Figure~\ref{fig:xct} shows the envelope function of the excitons
  in $k$-space for GaSe.
The first even $\sigma_h$ exciton forming the absorption peak "A"
  has an $s$-like shape, and is composed of the second-last
  valence band ($v_2$) to the first conduction band ($c_1$) transitions
  near $\Gamma$. %, as shown in Figure~\ref{fig:xct}.
The wavefunction of the excitons forming the "B" and "C" peaks
  are composed of the last valence band to the first
  conduction band transition.
They are located mostly on the saddle points along the $\Gamma-M$ line,
  and past the the point where the first two conduction bands cross.

In principle, these selection rules are exclusive to the single-layer structure,
  since they arise from the mirror symmetry of the layer,
  and the confinement of the wavevectors in the 2D Brillouin zone.
In the bulk counterpart of these systems,
  the wavefunctions could spread over several layers
  and relax the selection rules observed in the single layer.
However, experimental measurements show that this is not the case.
The strong anisotropy observed in the reflectivity
  of layered GaSe~\cite{liang_optical_1975}
  indicates that only a weak coupling exists between the layers,
  and that the selection rules of the single layer mostly hold
  in the bulk system as well.
We also note that the spin-orbit interaction (not included here)
  would allow for a mixing of valence band states with different
  parities \cite{LiSymmetrydistortedband2015a}.
Therefore, the $1s$ and $2s$ exciton states would realistically
  have a very small but finite optical response for in-plane polarization
  of the light.

\section{Conclusion}

In summary, we have computed the absorption spectra of single-layer GaSe
  and GaTe in the hexagonal phase.
Both of these materials exhibit sharp exciton peaks
  whose presence in the absorbance depend strongly
  on the angle of incidence and the polarization of the light.
We showed that the selection rules in these materials
  originates from the symmetry of the band states under in-plane
  mirror symmetry.
Non-vanishing optical transitions between the bands are either restricted to
  in-plane or out-of-plane light polarization
  which carry over to the excitons from those band pairs.
Our results explain the strong angular dependence of the absorption spectrum
  observed in few-layers GaSe \cite{liang_optical_1975}.
However, optical measurements on single-layer GaSe and GaTe
  have not been reported yet,
  likely due to the instability of these materials under
  ambiant air conditions\cite{BergeronOxidationdynamicsultrathin2017}.
The predicted absorption spectra reported in this work
  aim to guide future experiments.

\begin{acknowledgments}

This research was supported by the National Science Foundation
  under grant DMR-1508412 which provided for basic theory and formalism,
  and by the Center for Computational Study of Excited-State Phenomena
  in Energy Materials (C2SEPEM) funded by the U. S. Department of Energy,
  Office of Basic Energy Sciences, under Contract No. DE-AC02-05CH11231
  at Lawrence Berkeley National Laboratory which provided for algorithm
  and code developments and simulations.
The computational ressources were provided by the
  National Energy Research Scientific Computing Center (NERSC),
  a DOE Office of Science User Facility supported by the Office of Science
  of the U.S. Department of Energy
  under Contract No. DE- AC02-05CH11231,
  and the Extreme Science and Engineering Discovery Environment (XSEDE),
  which is supported by National Science Foundation
  Grant No. 787 ACI-1053575.

\end{acknowledgments}

\providecommand{\latin}[1]{#1}
\providecommand*\mcitethebibliography{\thebibliography}
\csname @ifundefined\endcsname{endmcitethebibliography}
  {\let\endmcitethebibliography\endthebibliography}{}

\end{document}